\documentclass[aps,twocolumn,showpacs,preprintnumbers,amsmath,amssymb]{revtex4}

\usepackage{graphicx}
\usepackage{dcolumn}
\usepackage{bm}
\newcommand{\be}{\begin{equation}}
\newcommand{\ee}{\end{equation}}
\newcommand{\bea}{\begin{eqnarray}}
\newcommand{\eea}{\end{eqnarray}}
\def\fct#1{\mathop{\rm #1}} 

\def\re{\fct{Re}}
\newcommand{\Eq}{Eq. \ref}
\newcommand{\Fig}{Fig. \ref}

\newcommand{\br}{r}
\newcommand{\bq}{q}
\newcommand{\bG}{G}
\newcommand{\bM}{M}

\newcommand{\bA}{A}
\newcommand{\bZ}{Z}

\def\<{\left\langle}                 
\def\>{\right\rangle}                 
\def\half{\frac{1}{2}}               
\begin{document}

\title{Thermodynamics and equilibrium structure of
Ne$_{38}$ cluster: Quantum Mechanics versus Classical}
\author{Cristian Predescu}
\affiliation{Chemistry Department, University of California at Berkeley, Berkeley, CA 94720}
\author{Pavel A. Frantsuzov and
Vladimir A. Mandelshtam}
\affiliation{Chemistry Department,
University of California at Irvine, Irvine, CA 92697, USA}

\date{\today}

\begin{abstract}
The equilibrium properties of classical LJ$_{38}$ versus quantum
Ne$_{38}$ Lennard-Jones clusters are investigated. The quantum simulations use both the
Path-Integral Monte-Carlo (PIMC) and the recently developed
Variational-Gaussian-Wavepacket Monte-Carlo (VGW-MC) methods. The PIMC
and the classical MC simulations are implemented in the parallel
tempering framework. The classical heat capacity $C_v(T)$ curve agrees
well with that of Neirotti {\it et al}  [J. Chem. Phys. \textbf{112},
  10340 (2000)], although a much larger confining sphere is used in
the present work. The classical $C_v(T)$ 
shows a peak at about 6 K, interpreted as a solid-liquid transition,
and a shoulder at $\sim$4 K, attributed to a solid-solid transition
involving structures from the global octahedral ($O_h$) minimum and
the main icosahedral ($C_{5v}$) minimum. 
The VGW method is used to locate and characterize the low energy states of
Ne$_{38}$, which are then further refined by PIMC calculations.
Unlike the classical case, the ground state of Ne$_{38}$ is a
liquid-like structure. Among the several 
liquid-like states with energies below the two symmetric
states ($O_h$ and $C_{5v}$), the lowest two exhibit strong
delocalization over basins associated with at least two classical
local minima.
Because the symmetric structures do not play an essential role in  
the thermodynamics of Ne$_{38}$, the quantum heat capacity is a
featureless curve indicative of the absence of any structural transformations. 
Good agreement between the two methods, VGW and PIMC, is obtained.
The present results are also consistent with the
predictions by Calvo {\it et al} [J. Chem. Phys. \textbf{114}, 7312
  (2001)] based on the Quantum Superposition Method within the
harmonic approximation. However, because of its approximate nature, the
latter method leads to an incorrect assignment of the Ne$_{38}$
ground state as well as to a significant underestimation of the heat capacity.

\end{abstract}

\maketitle


\bigskip
\section{Introduction.}

In this paper, we investigate equilibrium properties of the
Ne$_{38}$ Lennard-Jones (LJ) cluster. Particularly, we are interested in
how the equilibrium structure, energy, and  heat capacity as functions of temperature 
are affected by the quantum nature of the system.
Our interest is partly motivated by
recent advances in the development of  accurate numerical
Quantum Statistical Mechanics techniques as well as by their 
successful applications
to smaller Lennard-Jones clusters. As such, Refs.~\cite{HCest} report
fully converged heat capacity curves obtained by the Path-Integral
Monte-Carlo (PIMC) method for various Ar$_{13-n}$Ne$_n$ Lennard-Jones 
clusters. The $\mathrm{LJ}_{13}$ cluster is one of the smallest 
that exhibits a pronounced liquid-solid-like structural transition, according to the existence of a
peak in the heat capacity curve $C_v(T)$. On the other hand,
$\mathrm{Ne}_{13}$  is perhaps the largest Ne cluster treated so far
quantum mechanically on the entire range of interesting temperatures 
\cite{HCest, Examples_Ne13}. 


In the case of Ar or heavier rare gas clusters, the quantum effects
can  safely be treated as small perturbations \cite{C_T}. However, it
has long been known that the Ne clusters exhibit strong quantum
effects, not only in the low temperature regime, but also in liquid
phase \cite{Old_Ne}. Still, perhaps because $7$ and $13$ are magic
numbers, the quantum effects do not essentially change the structural
or thermodynamic properties of the respective Ne clusters. For
example, although the  heat capacity $C_v(T)$ around the
``solid-liquid'' transition temperature ($T\sim$ 10 K)  is reduced in
magnitude and shifted about $10\%$ toward lower temperatures as
compared with the purely classical LJ$_{13}$,  the heat capacity curve
for Ne$_{13}$ has a  behavior similar to that of LJ$_{13}$. It would certainly be
interesting to verify whether the quantum effects lead or not to
qualitative differences for larger Ne clusters, especially those that
do not have a magic number of particles. Of particular interest is the
$\mathrm{Ne}_{38}$ cluster, for which there is the additional question
of whether or not the quantum ground state is still localized in the
celebrated octahedral basin, a basin that contains the classical
global minimum \cite{MinLJ38} and which represents a  deviation from
the icosahedral Mackay packing characteristic of all the other neighboring
clusters \cite{Northby}.

An attempt to answer these questions for  a variety of rare gas LJ
clusters (including Ne$_{38}$) has been made by Calvo, Doye, and Wales
\cite{C_T}, using various approximations, particularly, the quantum
superposition method.  In this approximation, one treats all the
classical minima  of the potential energy surface independently and
within an harmonic approximation. Quite interestingly, the technique
predicts the absence  of the solid-liquid transition peak  in the heat
capacity curve for the quantum Ne$_{38}$ cluster. Also, it suggests
that the global minimum is no longer localized in the octahedral
basin. However, although many of the conclusions of Ref.~\cite{C_T}
remain qualitatively unchanged, we have found that the energy
estimates given by the harmonic approximation in this case are
unacceptably inaccurate. For instance, the harmonic approximation  
incorrectly assigns the ground state to a state from the
icosahedral basin (state $4$ from Table~\ref{table1}) that turns out
to have an energy about $24.6~k_\mathrm{B}\mathrm{K}$ greater than the
octahedral-based structure, invalidating the conclusions of
Ref.~\cite{C_T}. Nevertheless, we still find that  the configurations
of minimum quantum-mechanical  energy are indeed disordered,
liquid-like structures localized in the icosahedral basin.  Yet, it is
clear that techniques that are more sophisticated than the
Quantum Superposition Method must be utilized for a reliable study of the
thermodynamic and structural properties of the Ne
clusters.  As Calvo, Doye, and Wales acknowledge, it would be desirable
if their results could be verified by more accurate simulations,
although they note: ``It is very likely that simulating
$\mathrm{LJ}_{31}$ or $\mathrm{LJ}_{38}$ by quantum Monte Carlo
methods at thermal equilibrium is not practical with the current
computer technology.'' We attempt to give a partial answer to their
challenge by utilizing two different quantum algorithms, both of which
have been shown to accurately treat the smaller $\mathrm{Ne}_{13}$
cluster \cite{HCest, Neon13}.

The first algorithm is a random series  based  PIMC technique
\cite{RandomSeries, Pre04a, Pre04b} which, in principle, is an exact
method. It must be realized though that an accurate computation of the
heat capacity is a more demanding task for PIMC than the evaluation of
the average energy, for instance. Even for the latest heat capacity
estimators  \cite{HCest}, the standard deviation still grows as fast
as $T^{-2}$, when the temperature is lowered (the prediction is a
theoretical upper bound; in practice, we have always observed an
improved behavior, closer to $T^{-1}$).  At the same time, the heat
capacity itself decreases at a rate following perhaps a polynomial law
for a large range of low temperatures \cite{C_T} (of course, after
removing the part associated with the translational degrees of
freedom, the decrease is exponentially fast in the extreme low
temperature limit, due to the finite number of particles in the
cluster). Thus, the relative error increases rather severely at low
temperatures. Other limitations of the PIMC method come from the
increase in the numerical effort associated with the larger number of
path variables needed for low temperatures as well as from the
sampling difficulties related to the larger number of path
variables. However, the latter kind of limitations are surmountable,
either by employing path integral techniques having faster asymptotic
convergence \cite{Pre04a, PIMC_fast}  or by employing better sampling
strategies \cite{Pre04b, PIMC_sampling}. Most likely, the biggest gain
will come from the design of better heat capacity estimators. To cope
with ergodicity problems, the path integral technique has been
utilized in conjunction with the parallel tempering procedure
\cite{paral_temp},  in which a number of Metropolis walks at different
temperatures are utilized, with the configurations generated by walks
with similar temperatures swapped periodically.

In Refs.~\cite{gauss_CPL, Neon13} a novel method (VGW-MC) using
variational  Gaussian wavepackets in conjunction with a Monte Carlo
sampling of the initial conditions was developed and tested for
various benchmark systems including the Ne$_{13}$ Lennard-Jones
cluster.  
The original Heller's idea of using the VGW's
for approximate solution of the time-dependent Schr\"odinger equation 
\cite{Heller_JCP75,Heller_JCP76} 
was later followed
by many groups. Some
important, but certainly not complete references are
\cite{frozen_gauss,Metiu_many_gauss,metiu_dens,Coalson,G-MCTDH,Coalson_fly,QMD_review,Martinez,Buch}.
While most of the developments  were concerned with  
the real-time dynamics, the work of Metiu and
co-workers \cite{metiu_dens}, where the VGW's were adapted to the 
solution of the ``imaginary-time'' Schr\"odinger equations, 
most closely relates to the present VGW-MC method.

The VGW-MC is a manifestly approximate quantum statistical
mechanics approach, which is supposed to be exact only in the high
temperature limit or for a purely harmonic system. Quite
surprisingly, the results obtained by this method for general strongly
unharmonic systems turned out to be very accurate, even at low
temperatures. For example, in the case of the Ne$_{13}$ cluster,
nearly quantitative agreement with the existing PIMC calculations  was
achieved for the  heat capacity and equilibrium structural
properties. Although the method does employ Monte Carlo sampling, its
convergence properties at low temperatures are completely different
from those in the parallel tempering  Monte Carlo schemes: the initial
conditions for the Gaussian wavepackets are sampled by a primitive
Metropolis walk running at an inverse temperature  $\beta=\beta_{\rm
MC}$ sufficiently small to ensure ergodicity.  Each Gaussian is then
propagated to the higher values of $\beta$  making its contribution to
the partition function.  Thus, a single Metropolis walk is used to
obtain results for the entire temperature range of interest, also
circumventing the quasi-ergodicity problems at low temperatures.  As
argued and demonstrated in Ref.~\cite{Neon13}, it is the quantum
nature of the Ne system which ensures the convergence of the method, a
convergence that, at first glance, may seem completely
counterintuitive. Some numerical aspects of the VGW-MC method still
remain to be  explored in the future. Those include the use of more
efficient sampling strategies. For instance, the method can be
implemented in the parallel tempering fashion if needed.  However,
since the numerical scheme, as described in Ref.~\cite{Neon13}, worked
for  the present case of the Ne$_{38}$ cluster, no attempts  to
optimize its performance were made here.  Thus, in the present study,
the computations are done using both  parallel tempering PIMC and
VGW-MC.  The latter approach seems to have better convergence
properties at low temperatures, while the former is manifestly exact
(when the statistical errors can be made sufficiently small).

The double-funnel topology of the potential energy surface of the
$\mathrm{LJ}_{38}$ cluster \cite{Ne38_double-funnel}  has proved a
tough simulation challenge for most of the Monte Carlo
algorithms. This challenge has only recently been answered  by
Neirotti and coworkers \cite{Neirotti_Ne38}, who have employed the
parallel tempering algorithm to successfully simulate a
$\mathrm{LJ}_{38}$ cluster confined by a hard-wall potential with a
radius of $R_c = 2.25~\sigma_{\rm LJ}$. Extensive studies made by
Neirotti and coworkers  have shown that the parallel tempering
algorithm in the canonical ensemble cannot equilibrate the
$\mathrm{LJ}_{38}$ cluster in about $100$ million passes through the
configuration space, if the original Lee, Barker, and Abraham
\cite{conf_R} confining radius of $R_c = 3.612\; \sigma_{\rm LJ}$ is
utilized. However, they have argued that the thermodynamics of the
cluster remains basically unchanged if a smaller confining radius of
$R_c = 2.25\; \sigma_{\rm LJ}$ is utilized.

For the quantum $\mathrm{Ne}_{38}$ cluster  
the confining radius of $R_c = 2.25 \sigma_{\rm LJ}$ is
perhaps too small. For
instance, the radius of the configuration proposed in Ref.~\cite{C_T}
as a ground state for the $\mathrm{Ne}$ cluster (state 4 in Table 1)
is $R_c = 2.255 \sigma_{\rm LJ}$, after quenching to the closest
classical minimum. Clearly, such a state would have been made
inaccessible by the small confining radius utilized by Neirotti {\it
et al}. In the present work, we utilize a confining potential in the form of a steep polynomial, which is more suitable for quantum simulations. It has been argued that, for such an analytical form of the confining potential, a larger confining radius of at least $R_c = 2.65~\sigma_{\rm LJ}$ is necessary in order to ensure that the low-temperature thermodynamic properties of the classical cluster are not significantly altered \cite{Freeman_preprint}. 
For quantum systems, one needs to worry about the possibility that
the effects of quantum delocalization may lead to an even stronger influence
of the confining radius on the structural properties of the
$\mathrm{Ne}_{38}$ cluster. Thus, in the present study we have decided
to employ the original confining radius of $R_c = 3.612\; \sigma_{\rm
LJ}$. In making this choice, we have been also motivated by the
preliminary findings that the large quantum effects, through raising
the zero-point energies and barrier tunneling, radically reduce the
difficulty of the sampling problem. Yet, we had to make sure that the
thermodynamic and structural changes observed are solely due to
quantum effects and not to the larger confining radius employed. For
this reason, we have started our investigation with a Monte Carlo
simulation for the classical $\mathrm{LJ}_{38}$ cluster, which is
presented in the following section.

\section{Parallel tempering simulation of the $\mathrm{LJ}_{38}$ cluster}

As discussed in the Introduction, the first successful attempts
\cite{Neirotti_Ne38}  to compute the thermodynamic properties of the
$\mathrm{LJ}_{38}$ cluster by parallel tempering simulations have
revealed the difficulty in attaining ergodicity. The
$\mathrm{LJ}_{38}$ cluster has a double-funnel topology of the
potential energy surface \cite{Ne38_double-funnel}, with the
octahedral and icosahedral basins separated by a large-energy
barrier. The way parallel tempering \cite{paral_temp} tries to
overcome the high-energy barrier is by coupling a set of statistically
independent Monte Carlo Markov chains via exchanges of configurations
between replicas of slightly different temperatures. The swapping
events by themselves do not lead to new configurations. Rather, the
configurations from the low-temperature replicas  climb the ladder of
temperatures and reach the high-temperature replicas, where they are
destroyed and replaced with energetically more favorable
configurations.

The rate of equilibration of the parallel-tempering simulation depends
on the ability of the high-temperature Metropolis walkers  to jump
between the icosahedral and the octahedral basins with a sufficiently
high frequency. Let us mention that,  although they have enough energy
to do so, the replicas of highest temperature must defeat the action
of the entropy, which favors the configurations from the icosahedral
basin or, beyond the melting point, the configurations that are
associated with the liquid phase. At high temperature, the number of
local minima that become thermodynamically accessible increases as
$\sim \exp(\alpha N)$  (not counting the permutationally symmetric
configurations), with the dimensionality $N$ of the cluster
\cite{local_minima}. Therefore, the ability of the high-temperature
replicas to find configurations in the octahedral basin (the basin
associated with the global minimum) is quite low and decreases further
if the vapor pressure of the cluster is decreased, for instance, by
increasing the confining radius. As such, the finding of Neirotti {\it
et al} \cite{Neirotti_Ne38}  that the system is hard to equilibrate
even in $100$ million passes must not necessarily come as a
surprise. 

The Lennard-Jones parameters utilized in this paper  are those
employed in previous  studies of Ne clusters \cite{HCest}:
$\epsilon_{\rm LJ}=35.6$ K and $\sigma_{\rm LJ}=2.749$ A. The mass of
Ne atom was assumed to be $m=20$ a.u. 
In both classical and PIMC simulations, we have
employed the same analytical expression for the confining potential,
in the form of the steep polynomial
\begin{equation}
\label{2.3}
V_c(\mathbf{r_i})=\epsilon_{\rm
LJ}\left({\|\mathbf{r_i}-\mathbf{R_{cm}}\|}/{R_c}\right)^{20}.
\end{equation}
Here, $\mathbf{R_{cm}}$ is the center of mass, whereas $R_c =
3.612~\sigma_{\rm LJ}$ is the confining radius. Recent studies
performed by Sabo, Freeman, and Doll \cite{Freeman_preprint} have
demonstrated that low-temperature thermodynamic properties of the
cluster may be sensitive not only to the exact value of the confining
radius, but also to the shape of the potential. Because the polynomial
potential is not sufficiently flat for small values of
$\|\mathbf{r_i}-\mathbf{R_{cm}}\|$ when compared to a hard-wall
potential, a radius of at least $R_c = 2.65~\sigma_{\rm LJ}$ must be
utilized in order to prevent the disappearance of the pre-melting
solid-solid phase change \cite{Freeman_preprint}. However, beyond this
radius, the vapor pressure depends only slightly on the exact value of
the confining radius, for a large range of such values. The
low-temperature thermodynamic properties and, to a good extent, the
properties of the liquid phase remain invariant to the exact value of
the confining radius. Nevertheless,  the confining radius of $R_c =
2.65~\sigma_{\rm LJ}$, which is a minimal requirement for a classical
cluster, is no longer satisfactory for the quantum $\mathrm{Ne}_{38}$
cluster, whence our decision to employ the significantly larger Lee,
Barker, and Abraham value of $R_c = 3.612~\sigma_{\rm LJ}$ appears
justified, although not computationally optimal in the sense of
Ref.~\cite{Freeman_preprint}.

\begin{figure}[!tbp] 
   \includegraphics[angle=270,width=8.0cm,clip=t]{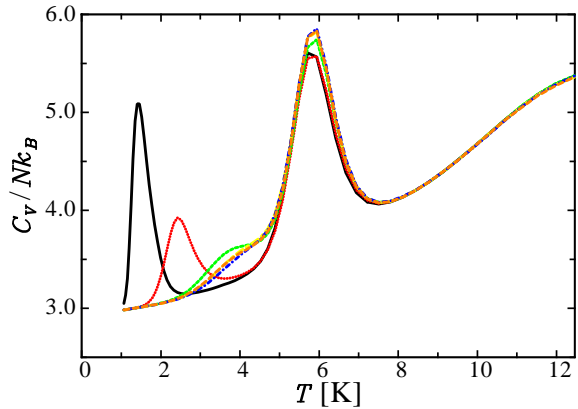}
 \caption[sqr] {\label{Fig:lj38pt}  Evolution of the heat capacity
profiles for the $\mathrm{LJ}_{38}$ cluster during the parallel
tempering Monte Carlo simulation. Each of the six curves has been
computed by collecting  averages for  successive groups of $50$
million passes. Early on in the simulation, the heat capacity shows a
fake maximum at low temperatures, maximum that is gradually washed out
and transformed in a ``shoulder.'' The last two curves are virtually
indistinguishable on the plot and the data utilized to compute them
have also been employed for the final result.}
\end{figure}

We have employed exactly the same parallel tempering strategy as the
one from Neirotti {\it et al} \cite{Neirotti_Ne38}, except for a
larger number of parallel replicas, $64$ instead of $40$, which have
been arranged in a geometric progression spanning the interval of
temperatures $[1.0~\mathrm{K}, 12.46~\mathrm{K}]$. As such, we do not
describe the technique here and, instead, refer the reader to the
cited reference. To begin with, we have performed a first simulation
in $100$ million sweeps (passes) through the configuration space,
which were preceded by $20$ million warming steps. The sweeps have
been divided in blocks of $5$ million each. Rather than using all $20$
blocks to compute the heat capacity, we have calculated two heat
capacity curves, each using $10$ blocks of data. This approach allows
us to evaluate the convergence of the simulation.  The two curves are
those from Fig.~\ref{Fig:lj38pt} that exhibit a clearly defined
low-temperature maximum, additional to the one associated with the
solid-liquid transition.

The appearance of the fake low-temperature peak in the heat capacity
curves early in the simulation tells us a lot about how parallel
tempering works. Such peaks are due to solid-solid transitions between
the configurations from the octahedral and the icosahedral basins,
transitions that lead to large energy fluctuations.
The transitions between the replicas are achieved through the replica
exchange mechanism. Because at the beginning of the simulation only a
few  octahedral configurations have been found, a low temperature
replica is forced to jump between octahedral and icosahedral
configurations through the exchange mechanism. As the simulation goes
on, more and more octahedral configurations are found. Because they
are energetically more stable, they are placed in the replicas of
lower energy. These low-temperature replicas are involved more and
more only in exchanges of configurations from the octahedral basin and
their energy fluctuations decrease: the fake heat capacity peak moves
to higher temperatures.

 \begin{figure}[!tbp] 
   \includegraphics[angle=0,width=8.0cm,clip=t]{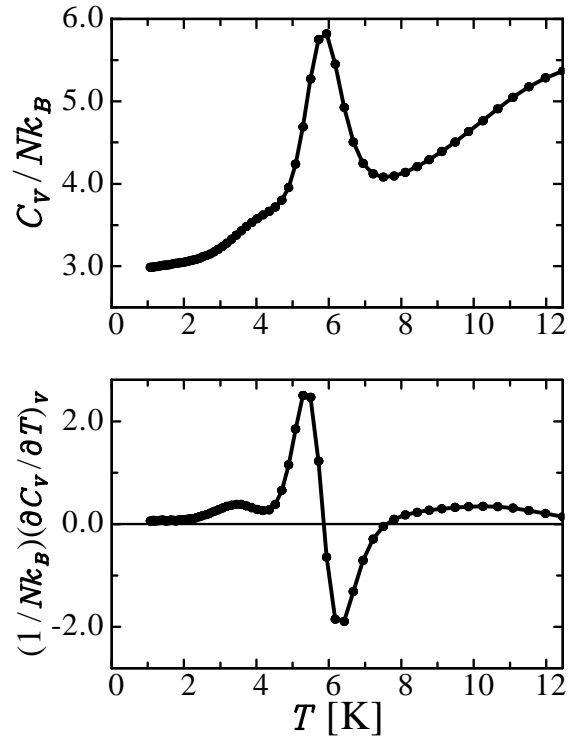}
 \caption[sqr] {\label{Fig:lj38ptp} Heat capacity per atom in
units of $k_B$ for the $\mathrm{LJ}_{38}$ cluster using a confining
radius of $R_c = 3.612\sigma_{\rm LJ}$. In the lower panel, the derivative
against temperature is given. As explained by Neirotti and coworkers
\cite{Neirotti_Ne38}, the small low temperature maximum in the derivative
curve is associated with a second order solid-solid phase transition
between the configurations from the octahedral and icosahedral basins.}
\end{figure} 

The mechanism described in the preceding paragraph also implies that
the rate of equilibration of the parallel tempering simulation can be
estimated from the rate at which the fake maxima move to the right.
More precisely, because the supply of newly found octahedral
configurations is steady, the peaks will move to the right roughly at
a constant rate until the heat capacity profile reaches the
equilibrium shape. Based on this line of reasoning, we have performed
two more simulations of $100$ million passes, each preceded by $20$
million warming sweeps, and each starting from the last configuration
of the previous simulation. The four heat capacity curves generated
using the new data are also shown in Fig.~\ref{Fig:lj38pt}, with the
last two being virtually indistinguishable. The data obtained in the
last $100$ million passes have been utilized to compute the heat
capacity curve from Fig.~\ref{Fig:lj38ptp}. Thus, the simulation has
needed about $260$ million passes for the equilibration phase and only
$100$ million for the accumulation phase.

The heat capacity curve obtained is almost identical to the one computed by
Neirotti and coworkers. Thus, our simulation
constitutes a direct proof of their assertion that the constraining
radius of $R_c = 2.25\sigma_{\rm LJ}$ is large enough not to essentially
change the low-temperature thermodynamics of the cluster. On the other
hand, our simulation reveals a very important property of the parallel
tempering algorithm. If the high-temperature replicas are capable of
finding the relevant configurations at a reasonable rate, then the
simulation will converge quite quickly and it should not
be stopped prematurely. Future research in the development of replica
exchange techniques should be concerned with designing ways to
identify which of the replicas generate the bottleneck configurations
and how one can improve the rate at which the configurations are
generated.

 \begin{figure}[!tbp] 
   \includegraphics[angle=0,width=8.0cm,clip=t]{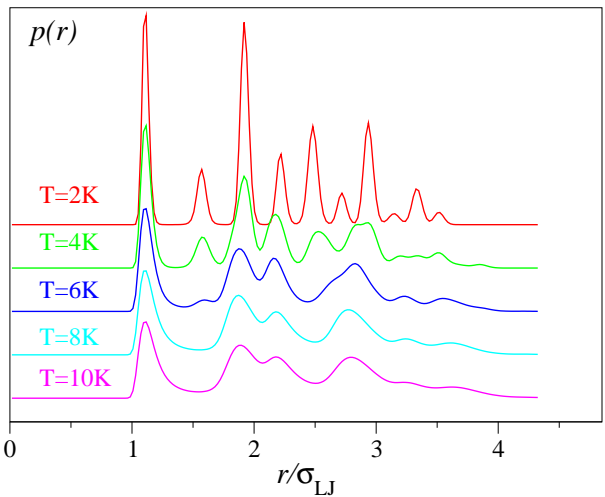}
 \caption[sqr] {\label{fig:pair_LJ38} Radial pair correlation
functions for the classical LJ$_{38}$ cluster. Notice the change in
the number and the position of peaks, as the temperature is lowered
from $T = 10~\mathrm{K}$ (liquid phase), to $T = 6~\mathrm{K}$ (mainly
structures from the icosahedral basin), and finally to $T =
2~\mathrm{K}$ (mainly structures from the octahedral basin). 
 }
\end{figure} 

In Fig.~\ref{fig:pair_LJ38}, we present the radial pair correlation
functions for different temperatures. The shape of the functions
changes  as the temperature is lowered from $T = 10~\mathrm{K}$, which
contains configurations characteristic of the liquid phase, to $T =
6~\mathrm{K}$, which mainly contains structures from the icosahedral
basin, and finally to $T = 2~\mathrm{K}$, which mainly contains
structures from the octahedral configurations. The association of the
shape of the radial pair correlation functions with the respective
configurations is based upon the analysis of several order parameters
performed by Neirotti and coworkers. (Also see below the comparison of
classical and quantum $p(r)$ of the same symmetries.)

\section{The Path Integral Monte Carlo technique}

It has been long recognized that the computation of  heat capacities
by path integral Monte Carlo  techniques is quite difficult, due to
the strong decrease in the heat capacity with the decrease in the
temperature. Our ability to report well-converged PIMC results for
relatively low temperatures (down to $1.78~\mathrm{K}$ for the
$\mathrm{Ne}_{38}$ cluster) relies on recent developments in the
design of direct path integral techniques (techniques that solely call
the potential function for the evaluation of physical
properties). Such developments include:  short-time approximations
having faster asymptotic convergence \cite{Pre04a}, more efficient
sampling techniques \cite{Pre04b}, and thermodynamic energy and heat
capacity estimators having lower variances \cite{RandomSeries, HCest}.
Because the PIMC technique has been extensively described in the cited
references, here, we only enumerate its salient features.

We employ a finite-dimensional approximation to the Feynman-Kac
formula in the form of a Lie-Trotter product  
\begin{eqnarray} \nonumber 
\label{eq:PIMC1}
\rho_n(x,x';\beta)=\int_{\mathbb{R}}d x_1 \ldots \int_{\mathbb{R}}d
x_n\; \rho_0\left(x,x_1;\frac{\beta}{n+1}\right)\nonumber \\ \ldots
\rho_0\left(x_n,x';\frac{\beta}{n+1}\right) 
\end{eqnarray} 
of a short-time approximation of the type
\begin{eqnarray}
\label{eq:PIMC2} \nonumber
\rho_0(x,x';\beta) = \rho_{fp}(x,x';\beta) \int_{\mathbb{R}} d
\mu(a_1) \cdots \int_{\mathbb{R}} d \mu(a_3) \\ \times
\exp\left\{-\beta \sum_{i = 1}^{4}w_i V\left[x_r(u_i) + \sigma
  \sum_{k = 1}^3 a_k \tilde{\Lambda}_k(u_i)\right]\right\}. 
\end{eqnarray}
The quadrature points $u_i$ and weights $w_i$ as well as the functions
$\tilde{\Lambda}_k(u)$ are designed such that the convergence 
\[
\rho_n(x,x';\beta) \to \rho(x,x';\beta)
\]
is as fast as $O(1/n^4)$. These parameters are universal, in the sense
that they are independent of the choice of potential $V(x)$, and are
given in Ref.~\cite{Pre04a}, reference that should be consulted for
further information. Here, we only mention that the short-time
approximation introduces additional path variables. The total number
of path variables for the diagonal elements of the $n$-th order
Lie-Trotter product is $4n + 4$, which is the number we report. The
number of evaluations of the potential function necessary to compute
the action for a particular path parameterized by the $4n + 4$ path
variables is also $4n + 4$. Therefore, the computational effort
relative to the number of path variables is the same as for the
trapezoidal Trotter approximation, yet the technique has a superior
asymptotic convergence, $O(1/n^4)$, as opposed to $O(1/n^2)$.

The Monte Carlo procedure is based on the fast sampling algorithm. As
observed in Ref.~\cite{Pre04b}, updating more than a few path
variables at a time in a Metropolis step results in a decrease in the
maximal displacements that is proportional to the square root of the
number of path variables updated. The effect is entropic in nature and
is roughly independent of the amount of correlation between the
variables that are sampled. On the other hand, if the variables are
updated separately, one needs to perform $4n + 4$ potential
evaluations for each of the $4n + 4$ variables. It has been argued
that the number of potential evaluations for an efficient update of
the path variables scales as $(4n+4)^2$, regardless of the strategy
utilized. The fast sampling algorithm is based on the observation
that, if $n = 2^k - 1$, then the $4n+4$ variables can be divided in $4
+ \log_2(n + 1)$ layers, with the path variables from each layer being
statistically independent. With a single evaluation of the action, one
can update all variables from a layer, independently. Therefore, the
cost to update all path variables in a statistically efficient manner
is $(n+1)[4 + \log_2(n + 1)]$, a  significantly better scaling.

Numerical experiments have shown that the number of path variables
necessary to achieve systematic errors comparable to the statistical
errors is $4n + 4 = 256$. The number of layers is $10$. A total of $10
\times 256 = 2560$ potential evaluations are needed  to efficiently
update each of the path variables associated with a given degree of
freedom. Similar to the Monte Carlo simulation for the classical
system, the path variables corresponding to different particles are
updated separately. We have randomly selected a particle and a layer
and updated the corresponding path variables. It follows that a pass
or sweep through the configuration space requires $38 \times 10 = 380$
elementary Metropolis steps. The simulation employed for the
computation of the heat capacity curve has consisted of $100$
accumulation blocks of $2$ thousand passes each. The accumulation
phase has been preceded by $50$ equilibration blocks. The Monte Carlo
replicas corresponding to different temperatures have been involved in
exchanges of configurations each $2$ passes, according to the parallel
tempering algorithm. We have employed a number of $64$ independent
replicas of temperatures arranged in a geometric progression spanning
the interval $[1.78~\mathrm{K}, 12.46~\mathrm{K}]$. The observed
acceptance rates for replica exchanges were larger than $40\%$.

The thermodynamic energy and heat capacity estimators are those
obtained by formal differentiation of the Lie-Trotter formula
\cite{RandomSeries}. The temperature differentiation can be performed
by a central finite-difference scheme requiring three points
\cite{HCest}. Thus, the overall computational effort for the quantum
simulation is a factor of $3 \times 2560 = 7680$ larger than for the
classical simulation, per Monte Carlo sweep. Fortunately, because of
the extensive quantum effects, one does not need so many passes as
for the classical simulation. In fact, as our later results show, the
configurations associated with the octahedral basin, which contains
the classical global minimum, have an unfavorable quantum energy. The
effective topology of the potential energy surface is drastically
simplified, with the octahedral basin being almost completely taken
out of the picture. Therefore, the ergodicity problems observed for
the classical simulation do not appear in the case of the quantum
simulation.

 In addition to the main Monte Carlo simulation, we have performed
 several Monte Carlo simulations at the fixed temperature of $T =
 1.78~\mathrm{K}$, in order to estimate the average energy of the
 configurations of minimal energy from Table~\ref{table1}. These
 simulations have utilized $4$ parallel streams of $50$ blocks each,
 for a total of $200$ accumulation blocks. The accumulation phase has
 been preceded by $25$ equilibration blocks. The simulations have been
 started from the centers of the Gaussian wavepackets of minimal
 energy that were obtained during the VGW simulation. We did not
 utilize parallel tempering for these simulations because  globally
 ergodic walkers would eventually leave the starting configurations
 and move to  entropically more favorable configurations. Only through
 broken  ergodicity can we  meaningfully associate the estimated
 energies with the input configurations. To verify the association, we
 have quenched several of the final positions of the simulated
 configurations. In all cases, we have recovered with a high
 probability (over $50\%$) the initial configurations. However, for
 both of the configurations $1$ and $2$, we have obtained another
 classical minimum, the basin of which was frequently visited. Thus,
 at least for the temperature of $1.78~\mathrm{K}$, the configurations
 of lowest energy are delocalized over basins associated with at least
 two classical minima.

\section{VGW-MC: Variational-Gaussian-Wavepacket Monte-Carlo.}

A general and detailed description of the VGW-MC method can be
found in Ref. \cite{Neon13}. 
In this section, we summarize those aspects of the technique that
regard the calculation of 
the partition function for a $N$-particle
system,
$$Z:=\mbox{Tr}\ \exp(-\beta \widehat H)$$
with  $\beta=1/k_{\rm B}T$.
The equilibrium energy is computed by differentiating the partition function
\be\label{eq:EZ}
E= k_{\rm B} T^2 \frac{\partial \mbox{ln}Z}{\partial T},
\ee
whereas the heat capacity is obtained by differentiating the energy
$$C_v=\frac{\partial E}{\partial T}.$$

In Cartesian coordinates, the Hamiltonian   is given by
\begin{equation}
\widehat H = -\frac{\hbar^2}{2} \nabla^{\rm T} \bM^{-1}
\nabla  + U(\br),
\label{Hamiltonian}
\end{equation}
with diagonal mass matrix $\bM=\mbox{diag}(m_i)$.
By $\br:=(\br_1,...,\br_N)^{\rm T}$, we define a $3N$-vector
containing the particle coordinates. 
$\nabla:=(\nabla_1,...,\nabla_N)^{\rm T}$ represents the gradient.

The partition function
is written as the integral
\be
\label{eq:GaussResol}
Z = \int d^{3N} \bq_0 \   K(\bq_0;\beta)
\ee
over the $3N$-dimensional configuration space,
where the integrand is
\be\label{eq:K}
K(\bq_0;\beta) := \langle\bq_0;\beta/2|\bq_0;\beta/2\rangle.
\ee
This expression is exact if the states $|\bq_0;\tau\rangle$ satisfy the
Bloch equation.
In the present framework, they
are approximated by the variational Gaussian wavepackets defined by
\bea\label{eq:Gauss}
\langle \br|\bq_0;\tau\rangle &:=&
  {(2\pi)^{-3N/2}\|\bG\|^{-1/2}} \\\nonumber &\times&\exp\left[ -\frac
    1 2 (\br-\bq)^{\rm T} 
\bG^{-1}(\br-\bq)+\gamma\right],
\eea
with the time-dependent parameters
$\bG=\bG(\tau)$, $\bq=\bq(\tau)$, and $\gamma=\gamma(\tau)$
corresponding, respectively, to the Gaussian width matrix (a $3N\times 3N$  real symmetric and
positive-definite matrix),
the Gaussian center (a real $3N$-vector),
and a real scale factor.
(Note the difference in the definition of the width matrix 
$\bG$ in \Eq{eq:Gauss} relative to its inverse  originally utilized in 
Ref. \cite{Neon13}).
Given the Gaussian approximation, the integrand in \Eq{eq:GaussResol} becomes
\be
\label{eq:K1}
K(\bq_0;\beta)  =
  (4\pi)^{-3N/2}\|\bG(\beta/2)\|^{-1/2}\ \exp [2\gamma(\beta/2)].
\ee

The Gaussian parameters are computed by solving
the system of ordinary differential equations
\be \label{eq:eqmotiond} \left\{
\begin{array}{l}
\dot{\bG}=  -\bG \<\nabla\nabla^{\rm T} U\> \bG + \hbar^2\bM^{-1}
\\
\dot {\bq} = -\bG \<\nabla U\>
\\
\dot \gamma = - \frac{1}{4} \mbox{Tr}\left(\<\nabla\nabla^{\rm T}U\> \bG\right)
 -  \<U\>
\end{array}\right.
\ee
starting from the initial conditions
\bea\label{eq:init0}
\bq(\tau_0) &=& \bq_0,
\\\nonumber
\bG(\tau_0) &=& \tau_0\hbar^2\bM^{-1},
\\\nonumber
\gamma(\tau_0) &=& -\tau_0 U(\bq_0),
\eea
which are defined for a sufficiently small but otherwise arbitrary
value of $\tau_0$. 
In \Eq{eq:eqmotiond}, $\<U\>$ represents the averaged (over the Gaussian)
potential, $\<\nabla U\>$, the averaged force and
$\<\nabla\nabla^{\rm T} U\>$, the averaged Hessian:
\bea\label{eq:U} \<U\>&:=& \nonumber
{\langle\bq_0;\tau|U|\bq_0;\tau\rangle K(\bq_0;2\tau)^{-1}},
\\
\<\nabla U\> &:=&
{\langle\bq_0;\tau|\nabla U|\bq_0;\tau\rangle K(\bq_0;2\tau)^{-1}},
\\\nonumber
\<\nabla\nabla^{\rm T} U\> &:=&
{\langle\bq_0;\tau|\nabla\nabla^{\rm T} U|\bq_0;\tau\rangle K(\bq_0;2\tau)^{-1}}.
\eea

Note the difference in the equations of motion (\ref{eq:eqmotiond}) 
relative to those originally derived  in Ref. \cite{Neon13}. Namely, 
the inverse of the matrix $\bG$ is not needed here and is not computed. 

For a potential with isotropic two-body interactions,
\be
U(\br)=\sum_{i>j} V(r_{ij}),
\ee
where $r_{ij}:=\br_i-\br_j$,
the Gaussian integrals in \Eq{eq:U}
are most conveniently evaluated by representing the pair potential
as a sum of Gaussians,
\be\label{eq:V}
V(r_{ij}) \approx \sum_{p=1}^P c_p \exp(-\alpha_p r_{ij}^2),
\ee
for certain parameters $c_p$ and $\alpha_p$ ($\re
\alpha_p>0$).
Simple potentials, such as the Lennard Jones potential, can be
accurately fit by only a few terms with real $\alpha_p$
\cite{Corbin,Neon13}. Here we utilize the same parameters as in ref. \cite{Neon13}.

Define the $3\times 3$ matrices:
\bea\nonumber &&
\bA_{ij} :=
\left(\bG_{ii}+\bG_{jj}-\bG_{ij}-\bG_{ji}\right)^{-1},
\\\nonumber &&
\bZ_{ij}(\alpha) := \alpha-\alpha^2(\alpha+\bA_{ij})^{-1},
\eea
where $\bG_{ij}$ denotes the corresponding $3\times 3$ block of the matrix $\bG$.
The analytic expression for the 3D Gaussian averaged over the variational
Gaussian wavepacket then reads
\be\label{eq:Vg0}
\<\exp\left(-\alpha r_{ij}^2\right)\>  = \sqrt{\frac{\|\bA_{ij}\|}{\|\bA_{ij}+\alpha\|}}
\exp\left[-\bq_{ij}^{\rm T}\bZ_{ij}(\alpha)\bq_{ij}\right],
\ee
where $\bq_{ij}:=\bq_i-\bq_j$. The  elements of the averaged gradient are
\bea\label{eq:Vg1} 
\<\nabla_k\exp(-\alpha r_{ij}^2)\>  = -2\< \exp(-\alpha r_{ij}^2)\>
\bZ_{ij}(\alpha)\bq_{ij},
\eea
for $k = i,j$, and  $\<\nabla_k\exp(-\alpha r_{ij}^2)\>=0$, for $k\ne i,j$. 
Finally, the four non-zero blocks of the second derivative matrix are
given by
\bea
&& \nonumber\<\nabla_i\nabla_i^{\rm T}\exp(-\alpha r_{ij}^2)\>
= \<\nabla_j\nabla_j^{\rm T}\exp(-\alpha r_{ij}^2)\>
\\ && \nonumber = -\<\nabla_i\nabla_j^{\rm T}\exp(-\alpha r_{ij}^2)\>
 = -\<\nabla_j\nabla_i^{\rm T}\exp(-\alpha r_{ij}^2)\>
\\ && \nonumber = 2\<\exp\left(-\alpha r_{ij}^2\right)\>
\left(2\bZ_{ij}(\alpha)\bq_{ij}\bq_{ij}^{\rm T}\bZ^{\rm
  T}_{ij}(\alpha)-\bZ^{\rm T}_{ij}(\alpha)\right). 
\\\label{eq:Vg2}
\eea

The most flexible form for the variational wavepacket is
the fully-coupled Gaussian (full matrix $\bG$). In this case, the numerical effort to solve the
equations of motion (\ref{eq:eqmotiond}) for the Gaussian parameters
scales as $N^3$. The extended 
acronym for the corresponding version of the method is FC-VGW.
A more approximate but computationally less intensive version (SP-VGW) employs 
single-particle variational Gaussian wavepackets
corresponding to a block-diagonal matrix $\bG$,  each block 
being a $3\times 3$ real
symmetric matrix representing a single particle. This results in $9N$
independent dynamical variables contained in the arrays $\bq$ and
$\bG$ and leads to the $\sim N^2$ numerical scaling, which is due to the
$\sim N^2$ terms in the potential energy.
Although both FC-VGW and SP-VGW are approximations, only the former gives exact 
results for general quadratic multidimensional potentials. 
Also, in the SP-VGW approximation, 
the motion of the center of mass is not separable. However,
as demonstrated in
Ref.~\cite{Neon13}, in the case of the Ne$_{13}$ Lennard-Jones
cluster,   the two methods give quite similar results for the heat
capacity, results that agree very well with those obtained by PIMC.

The integral in \Eq{eq:GaussResol} is most efficiently computed by the
Monte Carlo method.
The sampling strategy employed in the present work is as in Ref.~\cite{Neon13}.
The configurations sampled by a single Metropolis random walk  at a sufficiently
high temperature $T_{\rm MC}$ ($=1/k\beta_{\rm MC}$) are utilized to produce $Z=Z(T)$ for the entire
temperature interval of interest ($T<T_{\rm MC}$). That is, given the
sequence  $\{\bq_0^{(n)}\}\ \ (n=1,...N_{\rm MC})$ 
sampled according to
the probability distribution function $K(\bq_0^{(n)};\beta_{\rm MC})$,
the partition function for the temperature $T$ is computed with the help of the formula
\be
\label{eq:MC}
Z \approx \frac{1}{N_{\rm MC}}
   \sum_{n=1}^{N_{\rm MC}} \frac{
K(\bq_0^{(n)};\beta)}{K(\bq_0^{(n)};\beta_{\rm MC})}.
\ee

As extensively discussed and demonstrated in Ref.~\cite{Neon13}, for a
strongly quantum system as the Ne cluster, this
expression converges for all temperatures $T<T_{\rm MC}$.
This may seem to contradict the general experience with Monte Carlo simulations, as 
one expects the ensembles for different temperatures 
$T$ and $T_{\rm MC}$
to be quite different, fact that could potentially result in poor sampling. Despite this, 
\Eq{eq:MC} converges well. The explanation is that the entire Gaussian distribution,
which is broad at the high temperature $T_{\rm MC}$, shrinks when the
Gaussians are propagated to lower temperatures (the Gaussians fall
into the potential wells). Therefore, at all temperatures $T=1/k\beta$, the Gaussians
parameterized by $\bG(\beta/2)$, $\bq(\beta/2)$, and
$\gamma(\beta/2)$ are representative of the physically relevant region of
the configuration space.

\section{The ground state of $\mathrm{Ne}_{38}$ has liquid-like structure.}

The global  potential energy minimum of the LJ$_{38}$ cluster is a truncated
octahedron with energy $E_{cl}(O_h)=-162.943$ $k_{\rm B}$K. (Here and
throughout the paper the energy is reported as the energy-per-atom.)  The next-in-energy local minimum is
also a symmetric structure, namely an incomplete Mackay icosahedron
with $E_{cl}(C_{5v})=-162.310$ $k_{\rm B}$K. For the corresponding  quantum
system, it is natural to expect the ground state energy to be one of these symmetric
structures. On the other hand, the symmetric minima have the
stiffest potential, which, for sufficiently large values of the quantum delocalization 
parameter $\Lambda=\hbar/\sigma_{\rm LJ}\sqrt{m\epsilon_{\rm LJ}}$, may result
in high zero-point energies (ZPE). As such, the Harmonic  
Approximation (HA) \cite{C_T}
predicts that one of the disordered Mackay-based local potential minima that would
be assigned to a liquid-like structure in the classical simulations
(State 4 in Table~\ref{table1}) has the lowest ZPE.  
As argued in Ref. \cite{C_T}, this effect may also be accompanied
by the disappearance of the solid-liquid peak in the heat capacity
curve. This was confirmed by calculations using the Quantum
Superposition Method
for Ne$_{38}$, also within the HA.
Incidentally, more
accurate energy estimations (see below) show that the accuracy of the
HA is not sufficient to make a reliable prediction of the ground
state energy and structure. In fact, 
$E_{qm}(4)>E_{qm}(O_h),E_{qm}(C_{5v})$ (See Table~\ref{table1}), 
i.e., the state 4  has relatively high energy. 
Moreover, because of the strong delocalization of the 
eigenstates of Ne$_{38}$ over more than 
one classical minima (see below), the applicability of the
HA seems questionable.
However, from what
follows, the main qualitative conclusions of Ref.~\cite{C_T} remain correct.

In the present work, the procedure to search for the quantum ground state
consists of first generating a long classical
Metropolis walk at the temperature T=11.5~K (at which the random walk is ergodic). Every
once in  5000 classical MC steps, a configuration $\bq_0$ is selected
to set the initial conditions to propagate the variational  Gaussian wavepacket in
imaginary time to $\tau=1$ $(k_{\rm B}\mbox{K})^{-1}$, using the single-particle representation
(SP-VGW). 
During its propagation the energy of the VGW 
is computed using
\be
\varepsilon=-\half\frac{\partial} {\partial \tau} \mbox{ln}\<\bq_0;\tau|\bq_0;\tau\>.
\ee
Note that due to the variational nature of the Gaussian state, 
this is numerically identical to the  more familiar expression
$$\varepsilon=\frac{\langle\bq_0;\tau|\widehat H |\bq_0;\tau\rangle} {\<\bq_0;\tau|\bq_0;\tau\>}.$$
If the low-temperature state has sufficiently low energy it is 
further propagated  to $\tau=5$ $(k_{\rm B}\mbox{K})^{-1}$, at which point
the VGW becomes nearly stationary. If this state is  
distinguishable from all the previously identified 
low energy states, it is further refined by propagating it in imaginary time
again to $\tau=5$ $(k_{\rm B}\mbox{K})^{-1}$, but now using the more accurate
FC-VGW.

\begin{table}
\caption{
\label{table1}
Energies per atom in units of $k_{\rm B}$K of the six configurations
discussed in the text estimated by five different methods. 
The error bars for the PIMC results (twice the standard deviation) 
are all about $0.01 k_{\rm B}$K. In addition, the finite MC temperature, T=1.78 K, leads to 
a systematic state-independent shift of all the PIMC energies by about $0.1 k_{\rm B}$K.
States 1-4 have  
liquid-like structure with no particular symmetry. The classical
minima for states 1, 2 and 3 were obtained by quenching the quantum paths
during the PIMC simulations.
States 1 and 2 gave each at least two different
classical local minima. }
\begin{center}
\begin{tabular}{|c|c|c|c|c|c|} 
\hline
 State & FC-VGW & SP-VGW & PIMC & HA \cite{C_T} & Classical\\
\hline
$1$        & $-102.946$  & $-99.156$ & $-105.847$ & -          &
$\begin{array}{ll} -158.957 \\ -159.104 \end{array}$ \\\hline
$2$        & $-102.911$  & $-99.319$ & $-105.839$ & -          &
$\begin{array}{ll} -158.662 \\ -158.686 \end{array}$ \\\hline
$3$        & $-102.905$  & $-99.350$ & $-105.824$ & -          & $-158.312$ \\\hline
$4$ 	   & $-102.141$  & $-98.465$ & $-105.054$ & $-100.639$ & $-157.877$ \\\hline
$O_h$      & $-102.789$  & $-98.350$ & $-105.744$ & $-99.753$  & $-162.943$ \\\hline
$C_{5v}$   & $-102.814$  & $-98.682$ & $-105.700$ & -          & $-162.310$ \\\hline
\end{tabular}
\end{center}
\end{table}

%
%

\begin{table}
\caption{
\label{table2}
Same as Table~\ref{table1}
but with respect to the energy of state 1.}
\begin{center}
\begin{tabular}{|c|c|c|c|c|c} 
\hline
 State & FC-VGW & SP-VGW & PIMC \\
\hline
$1$         & 0     &  0     & 0     \\
\hline
$2$         & 0.035 & $-0.163$ & 0.008 \\
\hline
$3$         & 0.041 & $-0.194$ & 0.023 \\
\hline
$4$         & 0.805 &  0.691 & 0.794 \\
\hline
$O_h$       & 0.156 &  0.807 & 0.103 \\
\hline
$C_{5v}$    & 0.132 &  0.474 & 0.147 \\
\hline
\end{tabular}
\end{center}
\end{table}

It can be shown that in the $\tau\to \infty$ limit the Gaussian state
$|\bq_0;\tau\rangle$ becomes stationary. Its energy then gives an
upper estimate of the ground state energy. For small values of the
quantum delocalization parameter $\Lambda$ every minimum of the
potential energy results in its stationary Gaussian state. 
For large enough values of $\Lambda$, 
the quantum state may be delocalized over a number of local potential minima, which
is expected to be the case for the Ne
system. This significantly reduces the possible number of stationary states and, 
therefore, simplifies the search for the ground state
compared to the global optimization of the potential energy of the classical LJ$_{38}$ cluster.

In our calculations, a total of $10^6$ Gaussians have been generated. Out of those states, the three lowest 
energy states were selected for further analysis.
The results of our findings are summarized in Tables~\ref{table1} and~\ref{table2}. We also present the 
energy estimates for
state 4, which was incorrectly assigned to the ground state in Ref.~ \cite{C_T},
based on the harmonic approximation.
These four states appeared during the search many times with the hit rates
being $1.6\times 10^{-4}$,
$6.5\times 10^{-4}$, $2.8\times 10^{-4}$, and $10^{-5}$, for states 1,
2, 3, and 4, respectively. Neither octahedral ($O_h$) nor  icosahedral ($C_{5v}$) 
states were found during the search. As pointed out by Neirotti
\emph{et al} \cite{Neirotti_Ne38}, the fraction of highly-symmetric
structures having either $O_h$ or $C_{5v}$ symmetries is almost  zero
at the temperature of $11.5~\mathrm{K}$. Thus, there is the real
possibility that our simulation is not ergodic with respect to the
correct distribution of configurations at lower temperatures.

To address this issue, we have produced the $O_h$ and $C_{5v}$ states
by propagating the VGW starting from the corresponding classical
minima  and verified that their energies are not the lowest. In
addition, we have quenched  all the final configurations obtained
during the PIMC simulation, at the level of the VGW theory. These $64$
configurations, spanning the interval of temperatures
$[1.78~\mathrm{K}, 12.46~\mathrm{K}]$, have also produced the
configurations 2, 3, and 1 (in this order of abundance), as well
as some other configurations associated with the Metropolis walkers of
higher temperatures. The energies of the latter states are, however,
larger than those of states 1, 2, and 3. Quite
interestingly, state 1, which we believe to be the veritable
ground state, was not as frequently visited as state 2  during both the VGW and
PIMC simulations. Let us notice that the energies of these two states
are very close. Most likely, state 1 is not
entropically favorable on the interval of temperatures
$[1.78~\mathrm{K}, 12.46~\mathrm{K}]$. However, at even lower
temperatures, we may expect
state 1 to become the dominant species in the thermodynamic
ensemble. 
Unfortunately, our results so far (see the next section) are not capable to 
describe the low temperature regime accurately enough to state  
whether or not the transition from configuration 2 to 1 is capable of
producing a ``shoulder'' in the heat capacity curve.

\begin{figure}[!tbp] 
   \includegraphics[angle=0,width=8.0cm,clip=t]{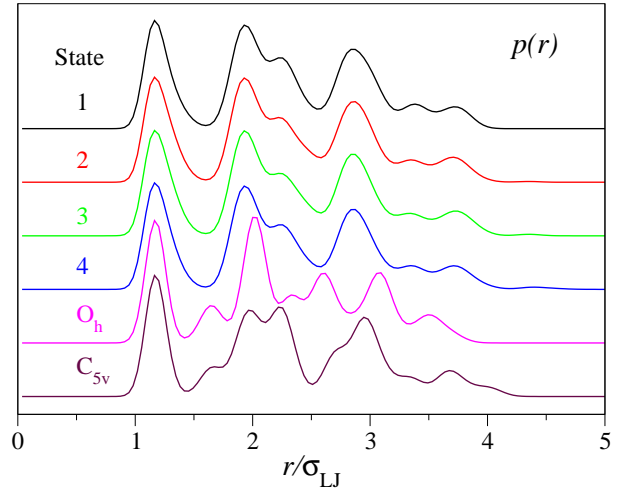}
 \caption[sqr] {\label{fig:paircor}    Radial pair correlation functions $p(r)$
for the six FC-VGW states from Table 1.  For the liquid-like states 
1-4 the radial distributions 
are nearly identical. 
}
\end{figure} 
 
In the context of the VGW approach, we found it sufficient and convenient
to characterize the structure of a stationary state by its radial pair correlation function,
\be
\label{eq:PairScal}
p(r):= \sum_{i>j}
\frac{\langle\bq_0;\tau|\delta(|r_{ij}|-r)|\bq_0;\tau\rangle}{\langle\bq_0;\tau|\bq_0;\tau\rangle}, \ee which is
computable with little numerical effort \cite{Neon13}.  The use of
angular-dependent distribution functions may be more  appropriate but
is more complicated. In \Fig{fig:paircor} we present $p(r)$ for the
six stationary states from Table~\ref{table1}. Quite interestingly,
for the liquid-like states 1-4, they are nearly identical. Moreover,
most stationary VGW states found in our calculations had the radial
distribution very similar to that of state 1, while only a few states
had $p(r)$ similar to that of the icosahedral state ($C_{5v}$). No
state with octahedral-like order was ever found. 

\begin{figure}[!tbp] 
   \includegraphics[angle=0,width=8.0cm,clip=t]{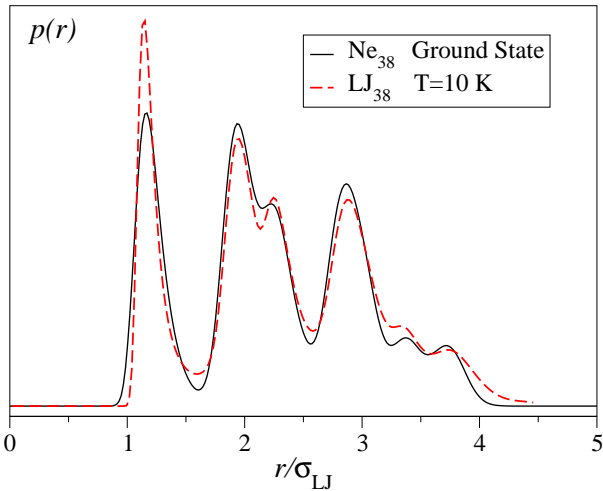}
 \caption[sqr] {\label{fig:paircomp} The radial pair correlation function $p(r)$
for the ground state of $\mathrm{Ne}_{38}$ 
and that for the
classical $\mathrm{LJ}_{38}$ cluster at $T=10$ K. 
The latter was scaled according to $p_{\rm scaled}(r)=\alpha p(r/\alpha)$ with $\alpha=0.97$
before plotting.}
\end{figure} 

Perhaps a more
convincing proof that the ground state of the quantum
$\mathrm{Ne}_{38}$ cluster corresponds to a disordered, liquid-like
state is the similarity  between the radial distribution function of
the quantum ground state and the radial distribution function of the
classical $\mathrm{LJ}_{38}$  system at the temperature of
$10~\mathrm{K}$ (see Fig.~\ref{fig:paircomp}).  The latter system is
in liquid phase at $10~\mathrm{K}$, as apparent from
Fig.~\ref{Fig:lj38ptp}. (Note that the classical radial correlation
function was scaled  using $p_{\rm scaled}(r)=\alpha p(r/\alpha)$ with
$\alpha=0.97$ in order to take into account the inflation of the quantum system 
relative to the classical one.)

Because it is more flexible than the SP-VGW,  the FC-VGW  provides a better
approximation for the ground state. While both approximations  fail to correctly describe the low energy rotation of
the cluster, the SP-VGW does not correctly describe the translational
motion. These are the obvious sources of the systematic errors in the
energy estimates using the Gaussian approximations. 
We found empirically that, for the FC-VGW, the main contribution to the 
error in the mean energy estimate E(T) is proportional to the system
size $N$ and is nearly temperature-independent.  
This explains the surprising agreement for
the heat capacity estimate $C_v(T):=\partial E(T)/\partial T$ computed
at the level of the FC-VGW theory and well-converged PIMC  
results for the Ne$_{13}$ cluster \cite{Neon13}. 

Qualitatively, the status of the SP-VGW based approximation is similar
to that for the
FC-VGW. However, the systematic error is about twice as big.  From
Tables~\ref{table1} and~\ref{table2}, we can see that  the energies
(per particle) of the selected states estimated  by   SP-VGW are
shifted by approximately 3 $k_{\rm   B}$K per atom relative to  those estimated
by the more accurate FC-VGW technique, independent  of the state. 
Also the SP-VGW gives different energy ordering for states 1, 2 and 3.

In order to verify the VGW results, we have also performed low temperature PIMC calculations 
using the initial conditions defined by the centroids of the corresponding Gaussians.
The simulation is explained in the last paragraph of Section~III and
it has been conducted at the temperature of $T=1.78$ K.  For each of
the six cases reported in Table~\ref{table1}, during the  course of
the MC simulation, the path was always localized in the  same
small region of the configuration space where it started. This was
checked by a quenching procedure which consisted in finding the
classical potential minimum nearest to the quantum path (see
Table~\ref{table1}). Note that each of the two lowest quantum states
(1 and 2) have resulted in two close minima upon quenching.  For each
of these two states we checked that the VGW gave the same  stationary
state when propagated from either of the two classical minima. 
For state 1 the two classical minima are separated by a barrier with
height by an order of magnitude smaller than the estimated value of
the ZPE, implying that the ground state must be delocalized over a
region including at least these two classical minima.

For the simulation
temperature  (1.78 K) utilized in the PIMC calculations,
the state energies are systematically overestimated by about $0.1$
k$_{\rm B}$K per atom. The latter error estimate was obtained by investigating
the temperature dependence of the corresponding VGW energies.

The discrepancy between the PIMC and the FC-VGW energies is about 3 $k_{\rm
  B}$K, independent of the state, which supports our previous
observation. This is clearly seen in Table~\ref{table2}, which gives the energies
with respect to the ground state energy. That is, after the
subtraction of the systematic errors, the agreement between the two
methods is quite remarkable.

\begin{figure}[!tbp] 
   \includegraphics[angle=0,width=8.0cm,clip=t]{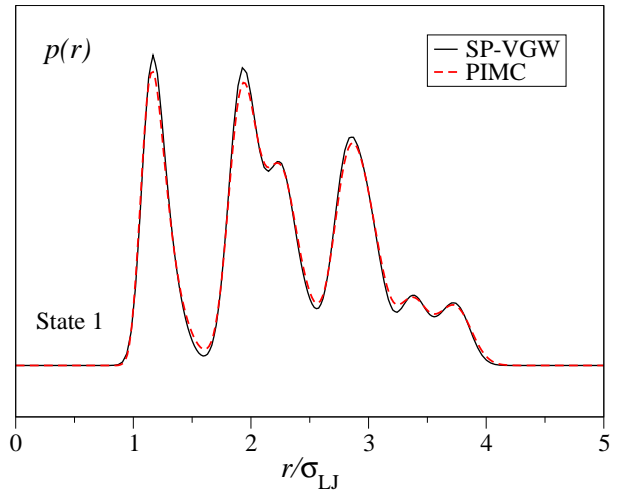}
 \caption[sqr] {\label{fig:comp_pair} The radial pair correlation function of the ground state
computed by the FC-VGW and the PIMC methods.
}
\end{figure} 

In \Fig{fig:comp_pair}, we compare the radial pair correlation function for state 1 computed by
FC-VGW and PIMC. The good agreement between the two results is another 
demonstration of the reliability of the FC-VGW method.

\section{Heat capacities for the quantum Ne$_{38}$ Lennard-Jones cluster.}

For the heat capacity calculations, we employed the less expensive 
SP-VGW version of the method, which  proved to be sufficiently
accurate for this purpose \cite{Neon13}. As for the classical simulations, the same confining
radius of $R_c = 3.612\sigma_{\rm LJ}$ is used here.

The results for the entire temperature range  (1 K $< T <$ 11.5 K) were obtained
within a single MC calculation using $T_{\rm MC}=11.5$ K.
The standard Metropolis algorithm was implemented using 25\% acceptance
rate. Every new Gaussian was sampled by randomly
shifting one of the atoms. Then this Gaussian, with the initial conditions
defined by \Eq{eq:init0}
at the initial inverse temperature
$\tau_0= 10^{-4} (k_{\rm B} \rm K)^{-1}$, was propagated up to
$\tau=1/2k_{\rm B}T_{\rm MC}$, where its acceptance probability was evaluated
and realized according to the Metropolis procedure.
Once a Gaussian wavepacket is accepted, it is further propagated up to
$\tau=0.5 (k_{\rm B} \rm K)^{-1}$, in order to cover the remaining
temperature range of interest. The total number of accepted  Gaussian wavepackets was $3\times 10^7$.
The calculations were performed on a 12-processor 1.4 GHz Opteron cluster. 
The cpu time on a single processor was about 0.5 seconds per accepted Gaussian wavepacket.

\begin{figure}[!tbp] 
   \includegraphics[angle=0,width=8.0cm,clip=t]{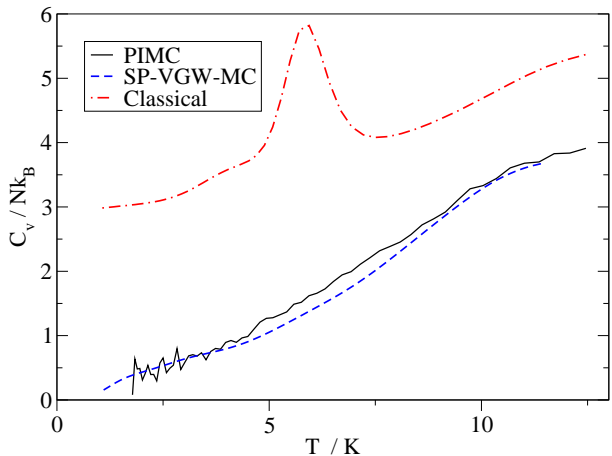}
 \caption[sqr] {\label{fig:CT} Heat capacity for the Ne$_{38}$
   Lennard-Jones cluster computed by two different 
methods. The result for a classical LJ$_{38}$ cluster is also shown. 
}
\end{figure} 

\begin{figure}[!hbp] 
   \includegraphics[angle=0,width=8.0cm,clip=t]{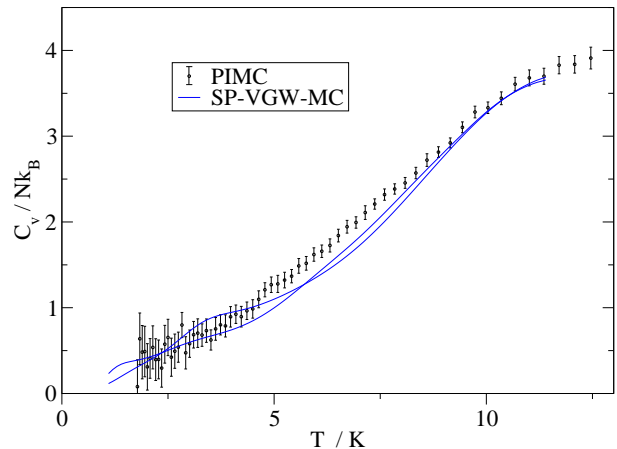}
 \caption[sqr] {\label{fig:CTerrors}Error bars for the results shown in \Fig{fig:CT}.
The two solid curves correspond to two independent VGW-MC calculations
each using $1.5\times 10^7$ MC steps.
}\end{figure}

The results for both the SP-VGW and PIMC simulations  are shown in
Figs. \ref{fig:CT} and \ref{fig:CTerrors}. The results for the latter
technique are for the interval $[1.78~\mathrm{K},
12.46~\mathrm{K}]$. At $1~\mathrm{K}$, the PIMC technique would have
required $512$ path variables and more parallel tempering replicas,
which we found rather expensive. The statistical errors for the SP-VGW
heat capacity were estimated by breaking the whole calculation into two
independent pieces consisting of $1.5\times 10^7$ MC steps each (see
\Fig{fig:CTerrors}).  Given the extreme complexity of the system,  the
agreement between the two methods is remarkable.  Within statistical
errors, one may safely conclude that there are no peaks in the caloric
curve of the quantum Ne$_{38}$ Lennard-Jones cluster for the
temperature  regime considered. Less clear is whether or not there is
a shoulder in the low-temperature portion of the heat capacity curve,
shoulder that could be assigned to a transition between 
configurations 1 and 2.

\begin{figure}[!tbp] 
   \includegraphics[angle=0,width=8.0cm,clip=t]{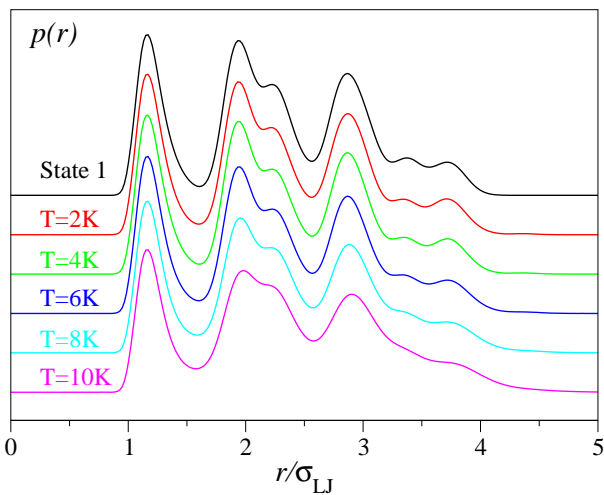}
 \caption[sqr] {\label{fig:pair_pimc}The radial pair correlation 
functions computed by the PIMC at various temperatures. 
The result for the ground state is also shown.}
\end{figure}

Finally, in \Fig{fig:pair_pimc}, we show the radial pair  correlation
functions computed by PIMC for several temperatures, including that of
state 1.  As opposed to what happens for the classical simulation (see
Fig.~\ref{fig:pair_LJ38}), the quantum results clearly indicate that
there is no abrupt change in the equilibrium structure of  Ne$_{38}$,
for the entire temperature interval considered. For all temperatures
considered, the quantum canonical ensemble for $\mathrm{Ne}_{38}$
consists mainly of configurations that Neirotti and coworkers
\cite{Neirotti_Ne38} have identified as pertaining to the liquid
phase. 

\section{Summary and conclusions}

We have investigated several thermodynamic and structural properties
of the quantum $\mathrm{Ne}_{38}$ cluster using two Monte Carlo
techniques: the variational Gaussian-wavepacket method and the path
integral method. As demonstrated by the results presented in the
preceding sections, the effective topology of the potential energy
surface is strongly affected by the quantum effects. For example, the
highly-symmetric octahedral and icosahedral configurations that
dominate the low-temperature classical canonical ensemble have
negligible contribution to the quantum canonical ensemble. The
$\mathrm{Ne}_{38}$ cluster is found to be essentially liquified for
all temperatures investigated.

The configurations of lowest quantum energy are
delocalized over  basins associated with at least two classical local
minima. This strong delocalization explains why the Harmonic
Approximation does not provide adequate estimates for quantum
energies. Moreover, perturbation theory based on inclusion of the
unharmonic terms would not improve the results of the latter. For these
techniques to be even applicable, the quantum states must be 
well localized around one classical minimum. The strong
delocalization is actually expected, given that $38$ is not a magic
number. The incomplete layer over the icosahedral core has high
mobility and is strongly affected by the quantum effects. As previous
simulations for $N = 7, 13$ have shown, for magic numbers, the quantum
effects, although strong, do not essentially change the shape of the
heat capacity curve.

An important conclusion of the present investigation is that the
quantum simulation techniques have matured enough that rather complex
system may be directly investigated. For the path integral technique
utilized, further work is necessary in order to provide better
estimates or criteria relating the quantum effects to the number of
path variables. Such criteria are important, for example, in the
context of parallel tempering. The present simulation has utilized a
number of $256$ path variables for the whole interval
$[1.78~\mathrm{K}, 12.46~\mathrm{K}]$. However, a number of $32$ path
variables would have been enough for the largest temperatures. The
computational resources saved would find a better utilization in
increasing the number of Monte Carlo steps for the higher temperatures
(at least by a factor of $8$). Criteria that could ensure a smooth as
well as adequate transition of the number of path variables with the
temperature are highly desirable.

The variational Gaussian wavepacket technique has demonstrated again
its reliability, providing results consistent with those obtained by
path integral simulations. Still, further work is necessary in order
to improve the convergence of the underlying Monte Carlo
simulation. Thus, the method must be adapted for use with parallel
tempering simulation techniques. Such an improvement would avoid
having to rely on the configurations generated by a single-temperature
Monte Carlo walk.

It is worth mentioning that the two quantum simulation techniques can
be used together as complementary tools, each having its particular
strengths. The path integral technique has the advantage of being
essentially exact, provided enough computational resources are
available. In the form it has been implemented in the present
application, it can be applied to the most general, many-body
potentials, for which the requirement of having analytic expressions for the Gaussian
integrals may not be practical. The VGW technique is generally
faster and provides results that are more amenable to
interpretation. It has also the advantage that it can be more easily
adapted for the study of quantum-dynamical properties, if
imaginary-time propagation of wavepackets is followed by real-time
propagation. As illustrated in the present application, the
VGW technique can be used to further
quench high-temperature path integral configurations to temperatures
so low that a direct path integral simulation is not feasible.

\begin{acknowledgments}
We are grateful to Jim Doll, David Freeman, Ken Jordan and Florent
Calvo for very useful discussions.
V.A.M. acknowledges the NSF support, grant CHE-0414110. He is an
Alfred P. Sloan research fellow. C.P. acknowledges support in part by
the National Science Foundation Grant Number CHE-0345280, the
Director, Office of Science, Office of Basic Energy Sciences, Chemical
Sciences, Geosciences, and Biosciences Division, U.S. Department of
Energy under Contract Number DE AC03-65SF00098, and the U.S.-Israel
Binational Science Foundation Award Number 2002170.
\end{acknowledgments}

\end{document}